\documentclass[preprint,aps,amsmath,superscriptaddress,nofootinbib]{revtex4-1}
\usepackage{graphicx}
\usepackage{appendix}
\usepackage{bm}
\usepackage{epsfig}
\usepackage{pstricks}
\usepackage{amsmath}
\usepackage{hyperref}
\usepackage{color}
\usepackage{verbatim}
\newcommand{\nn}{\nonumber}

\newcommand{\beq}{\begin{equation}}
\newcommand{\eeq}{\end{equation}}
\newcommand{\bea}{\begin{eqnarray}}
\newcommand{\eea}{\end{eqnarray}}
\newcommand{\OMIT}[1]{}

\definecolor{darkpurple}{rgb}{0.5,0,0.5}

\begin{document}
\preprint{ MIT-CTP 4457 }

\title{Classical Space-Times from the S Matrix }

\def\addCMU{Department of Physics, Carnegie Mellon University, Pittsburgh PA  15213, USA}
\def\addMIT{Center for Theoretical Physics, Massachusetts Institute of Technology, Cambridge, MA 02139, USA}

\author{Duff Neill}
\affiliation{\addMIT}

\author{Ira Z. Rothstein}
\affiliation{\addCMU}

\begin{abstract}
We show that classical space-times can be derived directly from the S-matrix for a theory of 
massive particles coupled to a massless spin two particle. As an explicit example we derive
the Schwarzchild space-time as a series in $G_N$.  At no point of the derivation is any
use made of the Einstein-Hilbert action or the Einstein equations. The intermediate steps
involve only on-shell S-matrix elements which are generated via BCFW recursion relations
and unitarity sewing techniques. The notion of a space-time metric  is only introduced at the end of
the calculation where it is extracted by matching the potential determined by the S-matrix
to the geodesic motion of a test particle. Other static space-times such as Kerr follow in a similar manner.
Furthermore, given that the procedure is action independent and depends only upon the
choice of the representation of the little group, solutions to Yang-Mills (YM) theory can be generated
in the same fashion. Moreover, the squaring relation between the YM and gravity three point functions
shows that the seeds that generate solutions in the two theories are algebraically related.
From a technical standpoint our methodology can also be utilized to calculate
quantities relevant for the binary inspiral problem more efficiently then the more traditional Feynman diagram approach.

\end{abstract}

\maketitle 

\section{Introduction}

It was pointed out long ago (\cite{Kraichnan},\cite{Gupta}, \cite{feynman}, \cite{weinberg}) that requiring the S-matrix of  
a massless spin two particle to be unitary and Lorentz invariant
places very strong constraints on the allowed gravitational dynamics.  Weinberg \cite {weinberg} gave an elegant
proof that these requirements lead to both the need to couple to a conserved stress energy
tensor as well as the equivalence principle. Using these simple assumptions one may side step Einsteins' geometric
construction completely to derive the Einstein-Hilbert action \cite{deser}  and  its associated set of
field equations.

Here we would like to  push this non-geometric program even further and ask, can we
generate classical space-times without going through the intermediate stages of
having to use the variational principle on a space-time action? Since the algorithm is action independent,  gravity is distinct from Yang-Mills 
theory only in its choice of representation of the Lorentz group.   As such we are able to show,  via the gauge-gravity ``squaring relation" \cite{GSW},    the
direct relation between classical solutions in the two theories. 

Additional motivation for this work stems from the fact that the methods of this paper   can
greatly simplify the calculations relevant for the binary inspiral problem. Recent progress in effective field
theory techniques \cite{NRGR} has shown that there may be some advantages to calculating
both potentials as well as radiative moments by considering an effective field theory of
sources coupled to gravitons whereby there is a manifest factorization of potential and
radiation degrees of freedom. In this methodology one calculates via Feynman diagrams
which we now know, is not the most economic way of organizing calculations.
The large gauge redundancy leads to monstrous intermediate results that
eventually collapse into simpler expressions. This is manifested 
in the 3PN \cite{foffa} calculation of the potential between non-spinning compact objects which
involves the calculation of one hundred diagram. This number jumps by an
order of magnitude at 4PN. Thus one would like to utilize the same techniques that have proven so successful
in simplifying calculations of Yang Mill scattering amplitudes \cite{ampreview}, to streamline the calculation of
these post-Newtonian potential and radiative moments.

At first site it may seem that this goal is not achievable since the aforementioned techniques  are typically applied to the 
calculation of  on-shell S-matrix elements, whereas we are interested to calculating
a classical potential, an off-shell quantity. As we discuss below this issue is resolved
by utilizing the unitary nature of the S-matrix to  extract
the relevant off-shell potential from the on-shell S-matrix element.

We construct the relevant gravitational scattering amplitudes through the  S-matrix bootstrap program, using the BCFW recursion \cite{BCFW} algorithm for tree amplitudes, and extending it into loop level calculations with unitarity methods \cite{BDDK}.  The potential is then determined by matching the full theory to
an effective theory, such that the potential is well defined and infra-red finite.
We calculate in a systematic expansion in the relative velocity, assuming a virialized orbit such that
$\frac{\vec{p}^{\,2}}{m^2}\sim\frac{G_N m}{r}\ll 1$, i.e. the post-Newtonian (PN) approximation.
Furthermore, we are only interested in the classical limit 
which  corresponds to a separate expansion in large angular momentum of the system: $|\vec{p}\times\vec{r}|\sim m|\vec{r}| \gg 1$, in units where $\hbar=1$\footnote{In terms of the Mandelstam variables, these requirements can be expressed as: $\frac{t}{s-(m_1+m_2)^2}\ll\frac{s-(m_1+m_2)^2}{s}\ll 1$.}.


\section{Building the S-matrix}
To build the S-matrix we start with the BCFW recursion \cite{BCFW} algorithm, relating on-shell n-point amplitudes to lower point on-shell amplitudes. The recursion relations follow from a particular complex deformation of two external momenta, explicitly leaving them on-shell. The deformation depends only on a single complex number, $z$. Then a contour integral over $z$ relates the physical amplitude (with no deformation) to a sum over pole terms and perhaps a contribution in the ultra-violet as $z \rightarrow \infty$. That is, 
\beq
\label{BCFW}
A_{phys}=A(0)=-\sum_\alpha Res \left( \frac {A(z)}{z} \right)_{z=\alpha}-A(\infty),
\eeq
where $A(z)$ is the deformed amplitude. Given that tree level amplitudes are meromorphic functions, the poles $z_\alpha$ correspond to on-shell intermediate states and the pole term can then be written as a product of on-shell lower point amplitudes ($A_L$ and $A_R$):
\begin{align}
 Res \left( \frac {A(z)}{z} \right)_{z=\alpha}&=\frac{A_LA_R}{P_\alpha^2-m^2},
\end{align}
where $P_\alpha$ and $m^2$  correspond to the momentum and mass, at zero deformation, flowing through the left amplitude into the right amplitude. The resulting pole  generates the correct physical propagator for the particle connecting the two amplitudes. 
 A theory is considered ``BCFW constructible''  if for any amplitude, there exists at least one  choice of momenta whose deformation has no contribution at infinity. That is, the amplitude is given completely by the pole terms. All proofs of BCFW constructibility rely  at some stage on an examination of the action of the theory to determine the UV behavior as $z\rightarrow\infty$. But as we will argue, to actually construct the classical gravitational metric, these UV terms are irrelevant. The part of the amplitude given by BCFW recursion always determines the classical potential\footnote{The actual amplitudes needed are in fact BCFW constructible, however, to show this, we must resort back to an action.}. 

For the purpose of building the Schwartzschild solution, we need the scattering amplitudes for two massive scalars radiating $n$ gravitons. To seed the recursion relations one needs the three point amplitudes to initiate the recursion process. These amplitudes vanish in Minkowski signature with real momenta, but  in the recursion relations one always works with analytically continued on-shell momenta. This allows for non-vanishing three point amplitudes. After fixing the external helicity states in the three point amplitudes, one can constrain their functional form by demanding the amplitudes be eigenfunctions of the little group generators for each external particle \cite{CachazoSmatrix}. For scattering one graviton off of a massive scalar, this yields up to an overall factor the amplitude (for positive helicity)
\beq
A_{GR}(1,2,3^+)=\frac{1}{2}\frac{\langle q 1\!\!\!\slash 3]^2}{\langle q 3 \rangle^2}\eeq
where $q$ is the arbitrary null four vector.  Here we are using spinor helicity notation (for details see \cite{dixon})  and all momenta are incoming. Similarly for YM theory, we have the color-ordered amplitude:
\beq
A_{YM}(1,2,3^+)= \frac{1}{\sqrt{2}}\frac{\langle q 1\!\!\!\slash 3]}{\langle q 3 \rangle}.
\eeq
In general one also needs the three gluon/graviton amplitude, but to the order we work in this paper, these are unnecessary. Notice that the two amplitudes satisfy the celebrated ``squaring'' relation:
\bea\label{KLT}
M_{GR}(1,2,3)&=& M_{YM}(1,2,3)M_{YM}(1,2,3). 
\eea
Thus the on-shell gravity three-point amplitude can be derived directly from the YM three point amplitude. This leads to the interesting interpretation of gravity as a YM theory with a kinematic ``gauge group'', as pursued in \cite{kinematic_algebra_I} and \cite{kinematic_algebra_II}. This squaring relation was also exhibited for classical solutions to GR and YM in \cite{Saotome:2012vy}. 

\begin{figure}[h!]
    \centering
    \includegraphics[width=7cm]{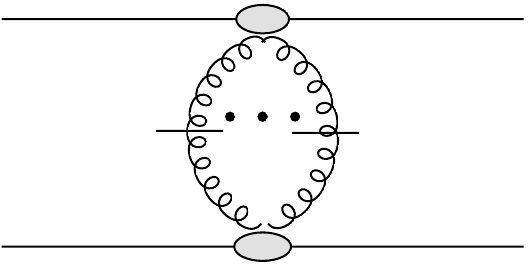}
\caption[1]{Reconstructing the full scalar-scalar  S-matrix by sewing together the
scalar-scalar n point on shell scattering amplitudes.}\label{fig1}
\end{figure}

Once we have the gravitational on-shell tree level scattering amplitudes we may construct the potential by sewing together these amplitudes using generalized unitarity \cite{BDDK}. This determines the GR amplitude for the scalar-scalar scattering. Since we are interested only in the long-distance classical pieces of the scattering amplitude we need only consider t-channel cuts as shown in figure (\ref{fig1}). 
Unitarity relates the discontinuity in an amplitude to a product of lower loop amplitudes, with a summation over physical states being exchanged between the amplitudes. In this way, one can construct an integrand that has the same analytic and singularity structure as the corresponding sum of feynman diagrams\footnote{It is precisely these singular terms that determine the long distance interactions. Thus we can ignore the effect of possible rational terms missed in the unitarity method.}. For the determination of the potential, only a restricted set of cuts need to be considered. In particular only two particle irreducible  diagrams can contribute to the classical  potential (see the appendix for a proof of this statement). Furthermore classicality also implies  that we need not consider  loops with only massless particles. Thus to fix the $G_N^n$ contribution to the classical potential, one only needs to consider the contribution from the product of $ss\rightarrow (n)g$  tree amplitudes.


\section{Definition of the potential}
The classical potential for extended sources can be  extracted by working within the confines  world-line effective theory\cite{,vokos,NRGR} where
the sources are treated classically. In this method the potential follows by calculating all two-particle irreducible diagrams and the classical and quantum pieces are easily distinguished. However,  this procedure relies upon a space-time action and thus will not suffice for
our purposes. We must choose a different route and instead, we consider the S-matrix elements in the non-relativistic limit, from which one can extract the classical potential \cite{japanesegroups}.

For our purposes we define the potential as being a matching coefficient of a two body operator in an effective non-relativistic theory \cite{PS} which corresponds to the post-Newtonian expansion in the gravitational case.  The potential seems like a peculiar matching coefficient since it depends upon external momenta. However, since the three momenta is parametrically larger then the non-relativistic energy it should be considered a UV scale in non-relativistic effective theories, and as such, can be thought of as labeling fields \cite{LMR}. The Lagrangian for the non-relativistic scalars in the center of mass frame is then written as
\bea\label{4_scalar_potential_theory}
L&=&
 -\sum_i\sum_{p,q}
 V_i(\vec q,\vec p)
 \phi^\dagger_{\vec p+\frac{1}{2}\vec q}(x)\phi_{\vec p-\frac{1}{2}\vec q}(x)
 \phi^\dagger_{-\vec p-\frac{1}{2}\vec q}(x)\phi_{-\vec p+\frac{1}{2}\vec q}(x).
\eea
 Since we are not interested in radiation this action holds for both YM theory (color indices are suppressed) as well as GR, since the ``gauge"" field (graviton or gluon) has been integrated out. Furthermore, note that as opposed to the Wilson loop definition of the potential, this definition is infra-red safe and allows for the inclusion of finite mass effects. One may be concerned that we have violated our credo of no actions, however, this action is not an action for gravity. It is used simply to define a potential. This four-scalar effective theory is simply the quantum mechanics of two bodies interacting via a potential, as such we have not technically passed to the classical limit. We finally achieve the classical limit by finding the potential of the classical hamiltonian:
\begin{align}
V_{i}^{cl}(\vec{r},\vec{p})&=\lim_{|\vec{r}\times\vec{p}|\rightarrow \infty}\int\frac{d^3\vec{q}}{(2\pi)^3}e^{-i\vec{q}\cdot\vec{r}}V_{i}(\vec{q},\vec{p})
\end{align}
 $V_i$  are a set of potentials each with a definite scaling in $v$, the relative velocity. In both GR and YM theory we will assume that the system of interest is in a virialized orbit such that $mv^2 \sim V$ and $v \sim (g^2,G_N M^2)/( M r)$ at leading order in YM and GR respectively. In calculating the potential from the full theory scattering amplitudes one generates contributions from three momentum regions \footnote{This assumes one is always working in dimensional regularization \cite{BS}.} as discussed in the appendix: soft, $k^\mu \sim (mv,m\vec v)$, potential, $k^\mu \sim (mv^2,m \vec v)$ and hard $k^\mu \sim (m,m)$. The latter region contributes only to the quantum mechanical part of the potential which is not of interest to us in this paper. The classical corrections to the Coulomb (Newton) potentials will be suppressed by powers of  $\frac{g^2}{mr} (G_NM/r)$ upon Fourier transformation.

It is sometimes stated that the classical potential is ambiguous. The potential itself is not a physical quantity, and as such, is not unique. There are two possible sources of ambiguity in the potential. The first ambiguity arises in the GR or YM theory, where traditionally the calculation must be performed by choosing a gauge for the graviton or gauge boson. However, at the level of the amplitude, this gauge dependence cancels \emph{if} one chooses physical external states. Indeed, in a bootstrapped S-matrix approach, one never has to choose a gauge, as we work always with gauge independent objects. Matching to the four-scalar theory, \eqref{4_scalar_potential_theory}, this will give rise to a unique potential in momentum space that is independent of the full theory gauge fixing. This potential is not more or less physical though than a potential we would have calculated had we matched off-shell in a specific gauge. Indeed, we see this from the second ambiguity which arises when we match the 4-scalar effective theory to the point particle hamiltonian in the classical limit. Here we must introduce the conjugate position to the relative momentum of the two bodies, and we are free to choose any other allowed coordinate system via canonical transformations, so long as the coordinate system remains asymptotically flat and retains a small relative velocity between the constituents of the binary system.

 Finally, there remains the fact that the two ambiguities can compensate for each other: had we matched off-shell in a particular gauge (inducing more terms in the potential), there exits a canonical transformation of the point particle Hamiltonian the would connect us to the on-shell matching, or vice versa. However, from the point of view of the sequence of effective field theories, these ambiguities are logically distinct, as they are allowed transformations in different theories.


When extracting $V$ from the matching calculation beyond leading order one must subtract from the full theory the contribution from inserting lower order potentials into time ordered products involving leading order interactions \footnote{The need for such  subtractions was first pointed out by Sucher in the context of the full relativistic field theory \cite{sucher}}. It is exactly this subtraction which leads to a potential which can be shown to be gauge equivalent (using transformations allowed in the theory \eqref{4_scalar_potential_theory}) to the classical source calculation. Since the process is iterative, before we can determine the proper subtractions we must first fix the lowest order potentials that arise from single gauge boson exchange.

\section{Tree-level Potential}
Extractions of  the gravitational potential from  S-matrix element has been performed in the past by calculating the Feynman diagrams generated by the Einstein-Hilbert action \cite{japanesegroups,bohr,Khrip,HR,HR1,HR2}. Here however, we are interested in generating potentials without any reference at all to the gravitational action. To accomplish this goal we  relate the potential, an inherently off-shell quantity, to on-shell scattering amplitudes. By doing so we bypass the need for any reference to an action, since tree level scattering amplitude may be generated simply from the knowledge of the primitive vertices (the three point scalar graviton and pure graviton amplitudes in particular) via the BCFW recursion algorithm. The leading order potential comes from ``sewing'' (not strictly in a unitarity sense) two scalar-scalar-graviton three point on shell amplitudes together. This would correspond to a BCFW shift of one of the scalar momenta on either side of the t-channel cut.

This four scalar tree level scattering process is strictly speaking not constructible via the BCFW methodology since when the external momentum is complexified the amplitude does not vanish in the limit where the complex momentum is taken to infinity. However, for our purposes this does not pose any obstruction since the  pole scales as $t \sim q^2$  thus any $z$ dependence in the numerator will lead to a quantum mechanical contribution.
 Thus ignoring this UV obstacle, fusing the leading order on-shell amplitudes generated from the three point seed amplitude, and  expanding in the relative velocity in  the center of mass frame, the  one graviton exchange  leads to the potential
\bea
\label{pot}
V(\vec q ,\vec{p})&=& -\frac{4\pi G_Nm_1 m_2}{ (\vec{q})^2}[1+\frac{\vec p^2}{2m_1m_2}\Big(\frac{3(m_1^2+m_2^2)+8 m_1 m_2}{2m_1m_2}\Big) 
\nn 
\\  &+&\vec p^4\Big(\frac{18 m_1^2 m_2^2-5(m_1^4+ m_2^4)}{8 m_1^3 m_2^3}\Big)
+...],
\eea
where we have used the relation $iM= -iV$. Where we have kept terms up to order $v^4$.
The order $v^4$ terms will be needed to extract the metric at order $G^2$, but are unecessary
if we are only interested in the 1PN potential.
There are additional corrections at this order  generated by loop diagrams. Since we wish to rely only on the basic three particle  S-matrix elements only we must  build up these diagrams with virtual lines by sewing together on-shell tree amplitudes to form  loop amplitudes with complexified  loop momenta . At  $O(v^2)$ we fuse together two four point amplitudes, which we generate via BCFW from the three point amplitudes. There are two possible (independent) helicity configurations for the gluons. The resulting four point functions are given by:
\bea
iA(1^+,2^+,3,4)&=&\frac{m^4}{4}\frac{[1 2]^2}{\langle 2 1 \rangle^2}\Big( \frac{1}{((1+3)^2-m^2)}+ \frac{1}{((1+4)^2-m^2)}\Big)   \nn \\
iA(1^+,2^-,3,4)&=&\frac{1}{4} \frac{\langle 2 ~\! 3\!\!\slash 1]^4}{(1+2)^2 }\Big( \frac{1}{((1+3)^2-m^2)}+ \frac{1}{((1+4)^2-m^2)}\Big).
\eea
All of the one loop diagrams can now be generated by sewing together the four point amplitudes in all possible helicity configurations where the  gravitons are exchanged in the t-channel.

\begin{figure}[h!]
    \centering
    \includegraphics[width=7.5cm]{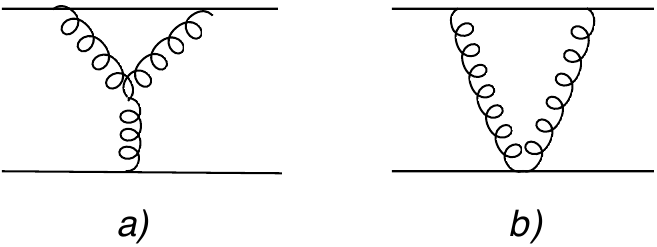}
\caption[1]{Graphs with the correct topology to contribute to the classical potential when asymptotically expanded about the potential region.}\label{Ygraph}
\end{figure}

 It is interesting to note that at this order, if we were calculating Feynman diagrams in a general gauge, the calculation would involve the three point graviton vertex as shown in (\ref{Ygraph}a). However, when generating the amplitude from generalized unitarity no information about the three graviton vertex has been utilized at this order.  The reason for this is that when we fuse the three graviton vertex with the scalar-graviton vertex, the t-channel graviton does not carry any of the complexified momentum and will thus not generate a pole in (\ref{BCFW}). This redundancy of the three graviton vertex  is consistent with the fact that when calculating the potential using classical sources as in \cite{NRGR} and  working with the Kaluza-Klein decomposition of the metric, one can show that there is no need for the three point vertex at 1PN \cite{KS}. Also note that sewing together two all-plus/minus  amplitudes does not contribute to the classical potential at this order as it is a pure box integral which has no classical contribution (see the appendix for a proof). 
\section{ The 1PN correction to the potential}
Given the cut-construction we now go about extracting the classical piece of the one loop potential at the level of the integrand. By the usual matching procedure in effective field theories  this involves calculating the
full theory graphs and subtracting the iteration of the tree level potential generated in the
effective theory. However, most of this work is unecessary as we are only interested in a small
subset of the full result. 

Instead we may determine which scalar integrals can contribute classical pieces using 
power counting arguments. We begin by noting that all classical pieces must scale as
$Gm^2/q^2 (G m q)^n$, with quantum corrections inducing multiplicative factors of 
$q/m $.  The claim is that the classical integrals   will always be topologically  two particle irreducible, as proven
in the appendix. Thus at one loop the only relevant scalar integrals will be triangles whose value is given by
\beq
I_{t}= \int \frac{d^4k}{(2\pi)^4} \frac{1}{k^2+ i\epsilon}\frac{1}{(k-q)^2+i\epsilon}\frac{1}{2m k_0+i\epsilon}= \frac{i}{16\pi^2 m^2}(1-\pi^2\frac{m}{2q}-\ln(q/m))
\eeq
The log and constant pieces are  quantum mechanical and can be dropped.

Extracting the classical piece and performing the momentum integrals yields the full theory result
\beq
iM= \frac{i(m_1+m_2)G_N^2\pi^2}{\mid\vec{q}\mid}\Bigg\{6m_1m_2+ \frac{9(m_1^2+m_2^2)+30m_1m_2}{2m_1m_2}\vec{p}^{\,2}+...\Bigg\}
\eeq
Note that we have included here the order $v^2 G^2$ contribution as well which is part of the 2PN correction to the potential. This piece will be necessary to calculate the metric at order $v^2$ as will be discussed below. To extract the classical potential at $O(v^2 G^2)$ we must subtract the $v^2$ corrections coming from the insertion of the sub-leading operators in the effective theory   into a time ordered products. There are two such insertions, shown in figures $3_a$ and $3_b$ which correspond to the insertion of a subleading potential and a kinetic energy correction respectively.

 The results for all of the possible time ordered products can be written as
 \beq
 M_A= \kappa_A \frac{G^2 \pi^2}{q}
 \eeq
 with
\bea
\kappa_a&=& 2 \frac{ m_1 m_2}{(m_1+m_2)}(3(m_1^2+m_2^2)+8 m_1 m_2) \nn \\
\kappa_b&=&  \frac{m_1m_2}{ (m_1+m_2)^2}\Big(\frac{1}{m_1^3}+\frac{1}{m_2^3}\Big) \nn \\
\kappa_c&=& 2  \vec p^2\frac{ m_1^2 m_2^2}{(m_1+m_2)^3} \nn \\
\kappa_d&=&   \frac{\vec p^2}{(m_1+m_2)m_1 m_2}(18 m_1^2 m_2^2-5 m_1^4-5 m_2^4)  \nn\\
\kappa_e&=&    \frac{\vec p^2}{(m_1+m_2)m_1 m_2}(3 (m_1^2+ m_2^2)+8 m_1 m_2))^2  \nn\\
\kappa_f&=&-\frac{3}{2}\vec p^2 \frac{m_1^4 m_2^4}{(m_1+m_2)^2}\left( \frac{1}{m_1^5}+\frac{1}{m_2^5}\right)  \nn \\
\kappa_g&=&\vec p^2 \frac{m_1^2 m_2^2}{q(m_1+m_2)^3}\left( \frac{m_1^3}{m_2^3}+\frac{m_2^3}{m_1^3}\right)  \nn \\
\kappa_h &=& 2\vec p^2 \frac{ m^2_1 m^2_2}{(m_1+m_2)^2}\Big(\frac{1}{m_1^3}+\frac{1}{m_2^3}\Big)(3(m_1^2+m_2^2)+8 m_1 m_2) .
\eea
Diagrams $c-h$ corresponds to $v^2G^2$ corrections which are required for the extraction of the 1PN metric extraction, but are unecessary for the calculation of the 1PN potential.

\begin{figure}[h!]
    \centering
    \includegraphics[width=8cm]{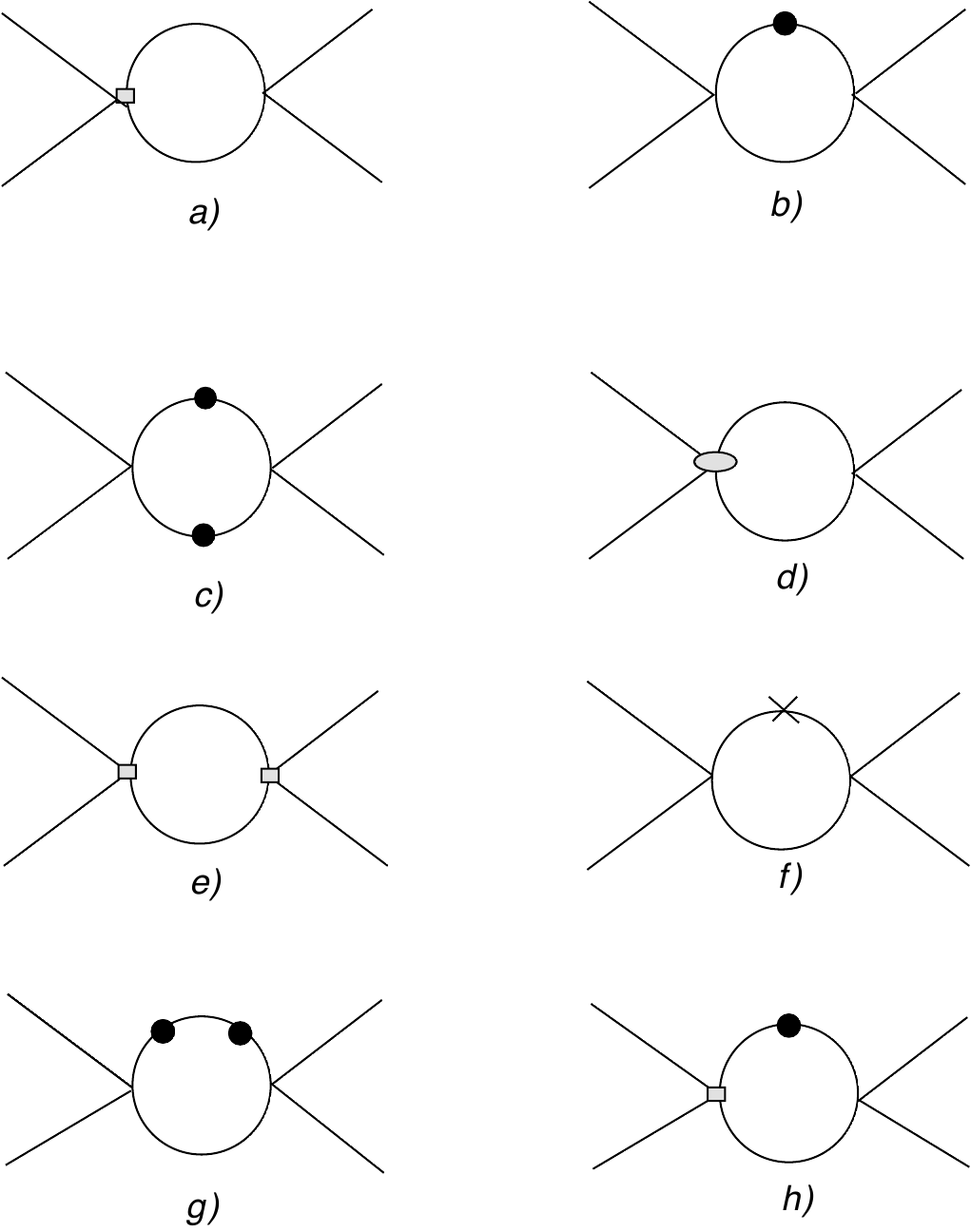}
\caption[1]{The time ordered product in the effective theory which must be subtracted from the full theory result. The square vertex is the order $v ^2$ Coulomb potential, while the dot corresponds to
the order $v^2$ kinetic term correction. The oval is the order $v^4$ Coulomb potential and the
cross is the order $v^4$ correction to the kinetic term.  Mirror image diagrams have been suppressed.
Diagrams $a$ and $b$ are 1PN while $c-h$ are 2PN and have been included only because we are  interested in the metric at order $G^2$.}\label{EFT}
\end{figure}
After subtracting the  sum of diagrams $a$ and $b$ from the full theory result and performing the Fourier transformation we are left with the potential
\bea
V&=&-\frac{G_N m_1 m_2}{r}\left(1+\frac{\vec p^2}{m_1m_2}\left(1+\frac{3(m_1+m_2)^2}{2m_1m_2}\right)\right)\nn\\
&+&\frac{G_N^2 m_1m_2(m_1+m_2)}{2 r^2} \left(1+\frac{m_1m_2}{(m_1+m_2)^2}\right)\nn\\
&+&\frac{G_N^2}{4r^2}\frac{\vec{p}^2}{m_1m_2}\Bigg(\frac{117m_1^2m^2_2+67(m_1m_2^3+m_1^3m_2)+10(m_1^4+m_2^4)}{m_1+m_2}\Bigg)
\label{EIH}
\eea
which agrees with the result in \cite{HR} and is gauge equivalent to the EIH potential.  In the probe ($m_1\ll m_2$) limit the resulting point-particle Hamiltonian is given by
\begin{align}
H(r,\vec{p})&=\frac{\vec{p}^{\,2}}{2m_1}-\frac{\vec{p}^{\,4}}{8m_1^3}+\frac{\vec{p}^{\,6}}{16m_1^5}-\frac{G_N m_1 m_2}{r}\left(1-\frac{G_Nm_2}{2r}+\frac{3\vec{p}^{\,2}}{2m_1^2}-\frac{5\vec{p}^{\,4}}{8m_1^4} -\frac{5G_N m_2\vec{p}^{\,2}}{2m_1^2r} \right).
\end{align}
corresponding to the potential,
\bea
V_{pr}&=&-\frac{G_N m_1 m_2}{r}\left(1-\frac{Gm_2}{2r}+\frac{3}{2} v^2-\frac{3G m_2}{2r} v^2\right).
\eea
\label{EIH}

\section{Extracting the metric by going to the probe limit}
Given the potential we may extract the metric order by order in powers of $v$ by comparing the potential to the world-line action. Assuming a static and isotropic solution  we may make the ansatz for the metric
\bea
g_{00}&=&1 + \sum _{i=1}^{\infty}A_{i}\lambda^i\nn \\
g_{0i}&=&0\nn \\
g_{ij}&=& -\delta_{ij}(1+\sum_{i=1}^\infty B_i \lambda^i)-\frac{r^ir^j}{r^2}(\sum_{i=1}^\infty C_i\lambda^i)  
\eea
where $A_i,B_i,C_i$ are constants and $\lambda=Gm_2/r\sim v^{2}$. The resulting potential is generated
from the world-line action (for the probe particle $m_1$) in the background of $m_2$
\beq
S=-m_1 \int dt \sqrt{g_{00}-v_iv_j g_{ij}}.
\eeq
This action can then be expanded and the potential is determined in terms of the set of unknown coefficients. Then by comparing with the result (\ref{EIH}) in the probe limit  we can read off the coefficients. At 1PN order we have $C_1=0$ and  $A_1=-2, A_2=2$ and $B_1=2$. Note that $C_2$ and $B_2$ do not contribute to the potential until 2PN. However we since we are only interested in the order $G^2$ piece of the solution we need not go to two loops to extract these coefficient, we only need to keep the order $v^2$ corrections to the full theory  one loop result and subtract all the order $v^2 G^2$ corrections in the effective theory. These 2PN graphs are  shown in figures $3c-h$ \footnote{The astute reader might be troubled by the fact that diagrams c-h have
include insertions of operators which scale as $v^4$, but only scale as $G^2 v^2$. The reason for
this is the singular nature of the Newtonian potential. To attain the more canonical power counting one could perform a rescaling.} . Subtracting these contribution we find that $C_2=0$ and $B_2=3/2$. Note that when working  in the center of mass frame all the $C_i$ will vanish, which corresponds to a particular choice of gauge. The final result for the metric is gauge equivalent to the more standard forms of the metric. Continuing in this way to higher orders we may build up the Schwarzschild solution.

\section{Discussion}

 This work can be generalized to construct the Kerr and Kerr-Newman solutions by considering the potential between a scalar and a charged higher spins  particles \cite{HR2}.
In addition, the  ideas presented here could be quite useful for the purpose of calculating higher order potentials in the PN expansion. The EFT techniques  described in \cite{NRGR} have been utilized to calculate potentials up to 3PN \cite{foffa}, and partial results have been reported at $4PN$ \cite{FS}. However, as one goes to higher orders in the PN expansion the number of diagrams begins to grow factorially. By using unitarity methods one avoids calculating Feynman diagrams in the full theory, and cumbersome intermediate stage expressions are avoided. It might seem however that this cost saving is not without drawbacks as one still has to calculate a multitude (albeit fewer) diagrams in the effective theory \footnote{Recall however that there is not need to go to 2PN to get the 1PN potential. We needed the 2PN result only for the purpose of extracting the metric.}. 
These diagrams are considerably simpler since they are in a scalar theory and the  topology  of the diagrams are pure bubbles.
Moreover, 
it may be possible to avoid the calculation of the EFT diagrams
altogether \cite{inprep}.

 Finally there is no obstruction to using these techniques to calculate higher order corrections to multipole moments.
Such multipole moments can be extracted by considering the same scattering processes discussed here, but now with
an additional graviton leg in the external state. This would involved matching onto a subsequent low energy theory
where the multipole moments are matching coefficients as discussed in \cite{NRGR, GR, PRR}.

\section{Acknowledgments}
We  thank I.W. Stewart  for useful  discussions. We also thank Zvi Bern for useful comments on the manuscript.  
We are especially indebted to Andreas Ross for discussions on unpublished work related to
\cite{HR1}. I.R. is supported  by DOE Grants DOE-ER-40682-143 and DEAC02-6CH03000. D.N. is supported by an MIT Pappalardo Fellowship and DOE Grant DE-FG02-05ER41360.

\appendix
\section{Proof of Two Particle Irreducible Nature of the Classical Contribution }

In this appendix we prove that only two massive particle irreducible diagrams (2PI), with one massive propagator per loop, can contribute to the classical potential. First we note that a simple dimensional analysis argument establishes the classical potential has the form:
\begin{align}\label{classical_scaling}
V^{cl}(q,\vec{p})&=\frac{Gm^2}{q^2}\Bigg(\sum_{n=0}^{\infty}(Gmq)^n\Big(\frac{\vec{p}}{m}\Big)^t\Bigg),
\end{align}
where $m$ stands for some generic mass. Quantum corrections will arise as deviations from this form, and $G$ acts as a loop counting
parameter.

We now will utilize the method of regions \cite{BS}, in which integrals are asymptotically expanded around the relevant regions
of momentum space. In our case there are only two regions, hard $k^\mu \sim m$ and soft $k^\mu \ll m$.
The hard region is easily seen to be purely quantum mechanical. At leading order in  $q/m$,  $q$ may be set to zero, and the
addition of any hard loop to a given diagram adds one factor of $G$, but to generate a classical correction the
extra loop integral must generate one, and only one, power of $q$. At leading order in $q/m$ the loop integral
is independent of $q$, while power corrections necessarily generate powers of $q^2/m^2$. Note the integrand will be independent of $\vec p \cdot \vec q$ in the center of mass frame and, as such, odd powers of $q$ will never be explicit in the numerator.
Thus we have established the fact that we need only consider loop momenta parametrically smaller then
$m$.  

\begin{figure}[h!]
    \centering
    \includegraphics[width=6cm]{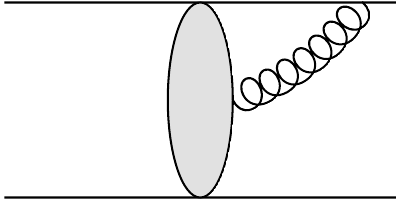}
\caption{The blob is a classical sub-loop.}\label{classical}
\end{figure}

Now we establish that 2PI diagrams with one massive propagator per loop contain a classical contribution.
The proof follows by induction. Suppose we have we take a graph which is 2PI and extract
its classical piece so that it scales as (\ref{classical_scaling}). Now we add an additional loop by connecting
a graviton from the body of the kernel to the scalar line as shown in figure (\ref{classical}).
The generic form of the resulting scalar integral is given by
\beq
I=G \int d^4k \frac{1}{(p+k)^2-m^2+i\epsilon} \frac{1}{k^2+i \epsilon} \frac{N(p^2,k^2,q^2,k \cdot q)}{P(k,q)}V(k,q,p),
\eeq
where $V$ has classical scalings and the factor of $P(k,q)$ scales as $k^2$, a result of the
additional propagator generated by hooking the extra graviton into the classical blob. $N$ is some polynomial of
its arguments.
The integral scales as 
\beq
I=G \int d^4k \frac{1}{(p+k)^2-m^2+i\epsilon} \frac{1}{k^2+i \epsilon} \frac{1}{P(k,q)}\sim S(N)/q^2,\eeq
where $S(N)$ is the scaling of the numerator, necessarily being  $\sim q^{2n}$. Thus in general we would
expect, generically, that  the resulting integral to contain both quantum and classical pieces.

As a technical aside we note that this soft region is usually broken up into two subregions which scale
as $k^\mu \sim (mv m\vec v)$ (soft) and $k^\mu \sim (mv^2,\vec v)$ (potential).
When considering the potential region, in a 2PI diagram, one is immediately faced with ill-defined integrals.
Considering the triangle topology in this region, we see that the energy integral is of the form $\int dk_0/k_0$.
However, this integral is cancelled by the zero-bin subtraction \cite{ZB} of the soft region, resulting in a well defined result.
This technical point is purely formal as it will not affect our proof, but merely establishes we need not consider the splitting of
the soft region further.

\begin{figure}[h!]
    \centering
    \includegraphics[width=6cm]{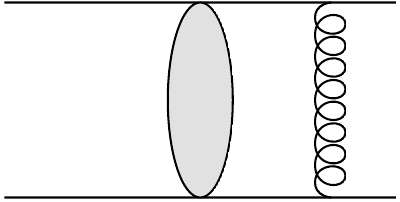}
\caption{A generic two particle irreducible contribution. The blob is a classical sub-loop.}\label{2PR}
\end{figure}

Next we examine the two particle reducible diagrams (2PR), that have two massive propagators in a loop. We will suppose that there is only one such loop. The generic form of such a contribution is shown in (\ref{2PR}), where the blob has classical scaling.
The resulting integral will be of the form
\beq
I = G \int [d^4k] \frac{1}{(k+q)^2}\frac{1}{(k+p)^2-m^2}\frac{1}{(p^\prime -k)^2-m^2} V(k,q,p) N(p^2,k^2,q^2,k \cdot q).
\eeq
In order for this integral to scale classically we would need 
\beq
I =  \int [d^4k] \frac{1}{(k+q)^2}\frac{1}{(k+p)^2-m^2}\frac{1}{(p^\prime -k)^2-m^2}  N(p^2,k^2,q^2,k \cdot q) \sim q,
\eeq
which is not possible, thus establishing our claim.

\end{document}